\documentclass[twocolumn,showpacs,preprintnumbers,amsmath,amssymb,superscriptaddress]{revtex4}

\usepackage{amsmath}
\usepackage{amssymb}
\usepackage{bm}
\usepackage{booktabs}
\usepackage{graphicx}
\usepackage{multirow}
\usepackage{listings}
\usepackage{xcolor}
\usepackage{textcomp}
\newcounter{MBQ}

\allowdisplaybreaks[1]

\def \refeq#1{(\ref{#1})}

\begin{document}

\preprint{
TUM-HEP-1094/17
}

\title{\boldmath 
  Enhanced electromagnetic correction to the rare $B$-meson 
  decay $B_{s,d} \to \mu^+ \mu^-$
}

\author{Martin Beneke}
\affiliation{
  Physik Department T31, 
  Technische Universit\"at M\"unchen,
  James Franck Stra\ss e~1,
  D -- 85748 Garching,
  Germany
}
\author{Christoph Bobeth}
\affiliation{
  Physik Department T31,
  Technische Universit\"at M\"unchen,
  James Franck Stra\ss e~1,
  D -- 85748 Garching,
  Germany
}
\affiliation{
  Excellence Cluster Universe,
  Technische Universit\"at M\"unchen,
  D -- 85748 Garching,
  Germany
}
\author{Robert Szafron}
\affiliation{
  Physik Department T31,
  Technische Universit\"at M\"unchen,
  James Franck Stra\ss e~1,
  D -- 85748 Garching,
  Germany
}

\date{August 25, 2017}

\begin{abstract}
\noindent
We investigate electromagnetic corrections to the rare 
$B$-meson leptonic decay $B_{s,d} \to \mu^+\mu^-$ from scales below the 
bottom-quark mass $m_b$. Contrary to QCD effects, which are entirely 
contained in the $B$-meson decay constant, we find that virtual 
photon exchange can probe the $B$-meson structure, resulting 
in a ``non-local annihilation'' effect. We find that this effect 
gives rise to a dynamical enhancement by a power of $m_b/\Lambda_{\rm QCD}$ 
and by large logarithms. The impact of this novel effect on the branching 
ratio of $B_{s,d}\to\mu^+\mu^-$ is about $1\%$, of the order of the 
previously estimated non-parametric theoretical uncertainty, and four 
times the size of previous estimates of next-to-leading order QED effects 
due to residual scale dependence. We update the Standard Model prediction 
to $\overline{{\cal B}}(B_s \to \mu^+\mu^-)_{\rm SM} 
  = (3.57 \pm 0.17) \cdot 10^{-9}$.
\end{abstract}


\maketitle
%


\noindent Rare leptonic decays $B_q\to \ell^+\ell^-$ of neutral $B$ mesons 
($q = d, s$ and $\ell = e, \mu, \tau$) provide important probes of 
flavour-changing neutral currents, since the decay rate in the 
Standard Model (SM) is predicted to be helicity- and loop-suppressed. 
Both suppressions can be lifted, for example, in models with extended Higgs 
sectors, in which case the leptonic decays constrain the scalar masses far 
above current direct search limits.

Only the muonic decay $B_s\to \mu^+\mu^-$ has been observed to 
date~\cite{Aaij:2013aka, Chatrchyan:2013bka}. The most recent measurement 
of the LHCb experiment for the untagged time-integrated branching ratio 
finds  $\overline{{\cal B}}(B_s \to \mu^+\mu^-)_{\rm LHCb} = 
(3.0^{+0.7}_{-0.6}) \cdot 10^{-9}$ \cite{Aaij:2017vad}, compatible 
with the SM prediction~\cite{Bobeth:2013uxa}
\begin{equation}
  \label{eq:SM:Br}
  \overline{{\cal B}}(B_s \to \mu^+\mu^-)_{\rm SM} 
  = (3.65 \pm 0.23) \cdot 10^{-9}\,.
\end{equation}
With higher experimental statistics and improvement in the 
knowledge of SM parameters, the accuracy of both results is expected 
to increase in the future, eventually providing one of the most 
important precision tests in flavour physics.

The neutral $B$-meson leptonic decays are indeed well suited for 
precision physics, because long-distance strong-interaction (QCD) effects, 
which cannot be computed with perturbative methods,  
are under exceptionally good control. This follows from the purely 
leptonic final state and the fact that the decay is caused by 
the effective local interaction 
\begin{equation}
  Q_{10} = \frac{\alpha_{\rm em}}{4\pi} \,\big(\bar{q} \gamma^\mu P_L b\big)
    \big(\bar{\ell} \gamma_\mu \gamma_5 \ell\big) \,, \quad
  P_L  \equiv \frac{1 - \gamma_5}{2}\,.
\end{equation}
The strong interaction effects are therefore confined to the 
matrix element
\begin{equation}
\langle 0|\bar{q} \gamma^\mu \gamma_5 b|\bar{B}_q(p)\rangle 
= i f_{B_q} p^\mu\,,
\label{fB}
\end{equation}
which defines the $B$-meson decay constant. $f_{B_q}$ can be computed 
non-perturbatively with few percent accuracy within the framework 
of lattice QCD~\cite{Aoki:2016frl}.

In this Letter, we report on an investigation of electromagnetic (QED) quantum
corrections to the leptonic decay which even at the one-loop order 
reveals a surprisingly complex pattern. As a consequence, the suppression 
of the correction due to the small electromagnetic coupling is partially 
compensated by a power-like enhancement in the ratio of the $B$-meson 
mass $m_B\approx 5~$GeV and the strong interaction scale $\Lambda_{\rm QCD} 
\approx 200~$MeV. While logarithmic enhancements due to collinear and 
soft radiation are well-known in QED and also appear in the 
process under consideration, the power-like enhancement arises due to 
a dynamical mechanism that to our knowledge has not been observed before.
A virtual photon exchanged between the final-state leptons and the 
light spectator antiquark $\bar q$ in the $\bar B_q$ meson effectively acts 
as a weak probe of the QCD structure of the $B$ meson. The scattering 
``smears out'' the spectator--$b$-quark annihilation over the 
distance $1/\sqrt{m_B\Lambda_{\rm QCD}}$ inside the $B$ meson, 
as opposed to the 
local annihilation through the axial-vector current in Eq.~(\ref{fB}). 
This provides power-enhancement and also shows that at first order 
in electromagnetic interactions, the strong interaction effects can no 
longer be parameterized by $f_{B_q}$ alone. Our calculation below 
shows that the effect is of the same order as the non-parametric theoretical 
uncertainty previously assumed to obtain Eq.~(\ref{eq:SM:Br}).

Before discussing the main result, we briefly review the computations and 
theoretical uncertainties entering Eq.~(\ref{eq:SM:Br}), referring to 
Ref.~\cite{Bobeth:2013uxa} for further details. The general framework employs 
the effective weak interaction Lagrangian, which generalizes the 
Fermi theory to the full SM, includes all short-distance quantum 
effects systematically by matching, and sums large logarithms between the 
scale $m_W$ of the $W$-boson mass and $\mu_b \sim m_b$ of the order of the 
bottom-quark mass, $m_b$. The SM prediction~\refeq{eq:SM:Br} includes 
next-to-leading order~(NLO) electroweak~(EW)~\cite{Bobeth:2013tba} and 
next-to-next-to-leading order QCD~\cite{Hermann:2013kca} corrections 
and the resummation of large logarithms $\ln(\mu_W/\mu_b)$ due to QCD
and QED radiative corrections by means of the renormalization-group (RG) 
evolution \cite{Bobeth:2003at, Huber:2005ig} down to $\mu_b$ at the same 
accuracy. Relevant to this work is the observation that unlike QCD effects, 
which are contained in $f_{B_q}$ to any order, QED corrections 
below the bottom mass scale $\mu_b$ have not been fully considered even 
at NLO.

The largest uncertainties in the SM prediction are of parametric origin:  
$4\%$ from the $B_s$ meson decay constant $f_{B_s}$, 
$4.3$\% from the quark-mixing 
element $V_{cb}$~\footnote{The determination of $V_{cb}$ from 
inclusive $b\to c\ell\bar\nu_\ell$ has been used.}, and  $1.6$\% from the 
top-quark mass. These uncertainties will reduce as lattice QCD calculations 
and measurements of SM parameters improve. Non-parametric uncertainties 
are due to the omission of higher-order corrections $\alpha_s^3, 
\alpha_{\rm em}^2, \alpha_s \alpha_{\rm em}$ in the QCD and QED
couplings $\alpha_s$ and $\alpha_{\rm em}$, respectively, and also 
$m_b^2/m_W^2$ from higher-dimension operators in the weak effective 
Lagrangian. Altogether, the non-parametric 
uncertainties have been estimated to be 
about $1.5$\%~\cite{Bobeth:2013uxa}. Among these, the renormalization scale 
dependence of $\overline{{\cal B}}(B_q \to \ell^+\ell^-)$
due to higher-order QED corrections accounts for only $0.3\%$. 
In view of such extraordinary precision, it is necessary to  
exclude the existence of unaccounted theoretical effects at the level of~1\%. 

Although NLO electromagnetic effects above the $b$-quark mass 
scale $\mu_b$ are completely included in Eq.~(\ref{eq:SM:Br}), this is not 
the case for photons with energy or 
virtuality below this scale. Since the decay involves 
electrically charged particles in the final state, only a suitably defined 
decay rate $\Gamma(B_q\to \ell^+\ell^-) + 
\Gamma(B_q\to \ell^+\ell^- + n \, \gamma )_{\rm cut}$ including photon 
radiation and virtual photon corrections is infrared finite and 
well-defined. Energetic photons are usually vetoed in the experiment and 
accordingly neglected on the theory side. Soft-photon emission from 
the final-state leptons is accounted for by experiments \cite{Aaij:2013aka, 
Chatrchyan:2013bka, Aaij:2017vad}. 
Initial-state soft radiation has been estimated to be very small based on 
heavy-hadron chiral perturbation theory~\cite{Aditya:2012im}.
The quoted measured branching fraction is corrected for soft emission 
and actually refers to the non-radiative branching 
ratio~\cite{Buras:2012ru}, as does Eq.~\refeq{eq:SM:Br}. 
For the purpose of the SM prediction~\cite{Bobeth:2013uxa} it was 
assumed that other NLO QED corrections below $\mu_b$ can not exceed the
natural size of $\alpha_{\rm em}/\pi \sim 0.3$\%. However, as we discuss 
now, the true size of so far neglected QED effects is substantially 
larger and in fact of the same order as the non-parametric theoretical 
uncertainty of 1.5\%. 

The primary challenge of NLO QED computations below $\mu_b$ consists in 
the reliable computation of non-local matrix elements. For example, a 
virtual photon connecting the spectator quark with one of the final-state 
leptons involves the QCD  matrix element
\begin{equation}
\label{eq:NLO:QED:ME}
\langle 0|\int d^4 x\,
T\{j_{\rm QED}(x), \mathcal{L}_{\Delta B=1}(0) \} 
|\bar{B}_q \rangle ,
\end{equation}
where $j_{\rm QED} = Q_q \bar q\gamma^\mu q$ is the electromagnetic 
quark current and $\mathcal{L}_{\Delta B=1}$ denotes the (QCD part of the) 
weak effective Lagrangian for $\Delta B = 1$ transitions. This matrix 
element bears close resemblance to the hadronic tensor that 
contains the strong-interaction physics of $B^+\to \ell^+\nu_\ell\gamma$ 
decay, which is known to be highly non-trivial (for example, 
Ref.~\cite{Beneke:2011nf}) despite its apparently purely non-hadronic 
final state.

In the following we focus on the muonic final state $\mu^+\mu^-$. We 
have analyzed the complete NLO electromagnetic corrections below the 
bottom mass scale $\mu_b$, counting the muon mass $m_\mu$ and 
spectator quark mass $m_q$ as $m_\mu\sim m_q \sim 
\Lambda_{\rm QCD} \ll m_b$ to organize the result in an expansion in 
$\Lambda_{\rm QCD}/m_b$. We then find that the electromagnetic 
correction to the decay amplitude is enhanced by one power of 
$m_B/\Lambda_{\rm QCD}$ compared to the pure-QCD amplitude. In the 
following we discuss only this formally dominant power-enhanced 
contribution, leaving the analysis of the complete QED correction 
to a separate publication. Note that the standard collinear and soft 
electromagnetic logarithms belong to these further, non power-enhanced 
terms, and are therefore not discussed here.

\begin{figure}
\begin{center}
\vspace{0.3cm}\hskip-4.5cm
\includegraphics[width=0.18\textwidth]{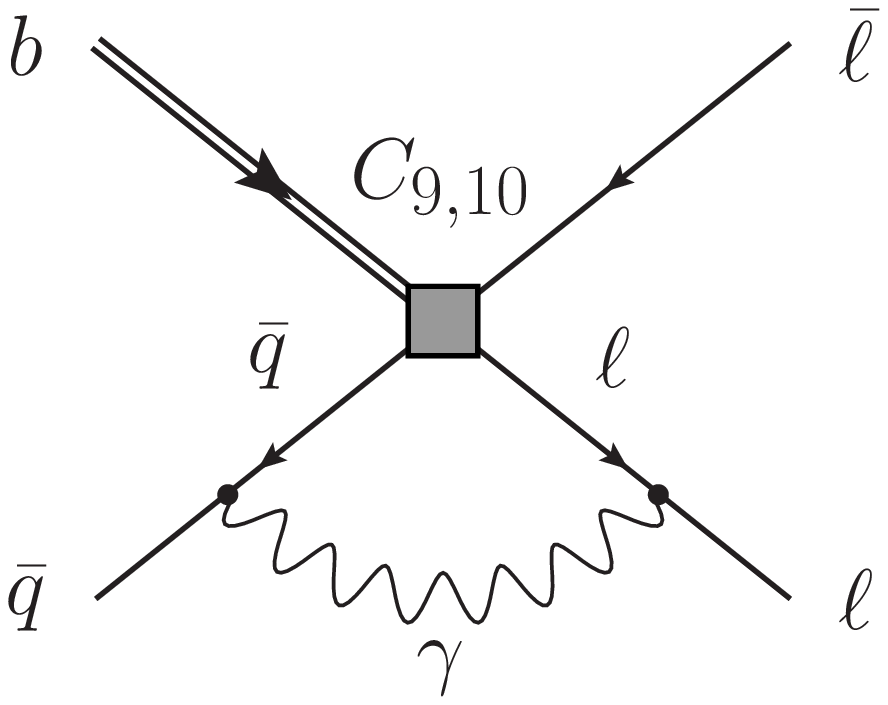}

\vspace{-2.35cm}\hskip3.7cm
\includegraphics[width=0.25\textwidth]{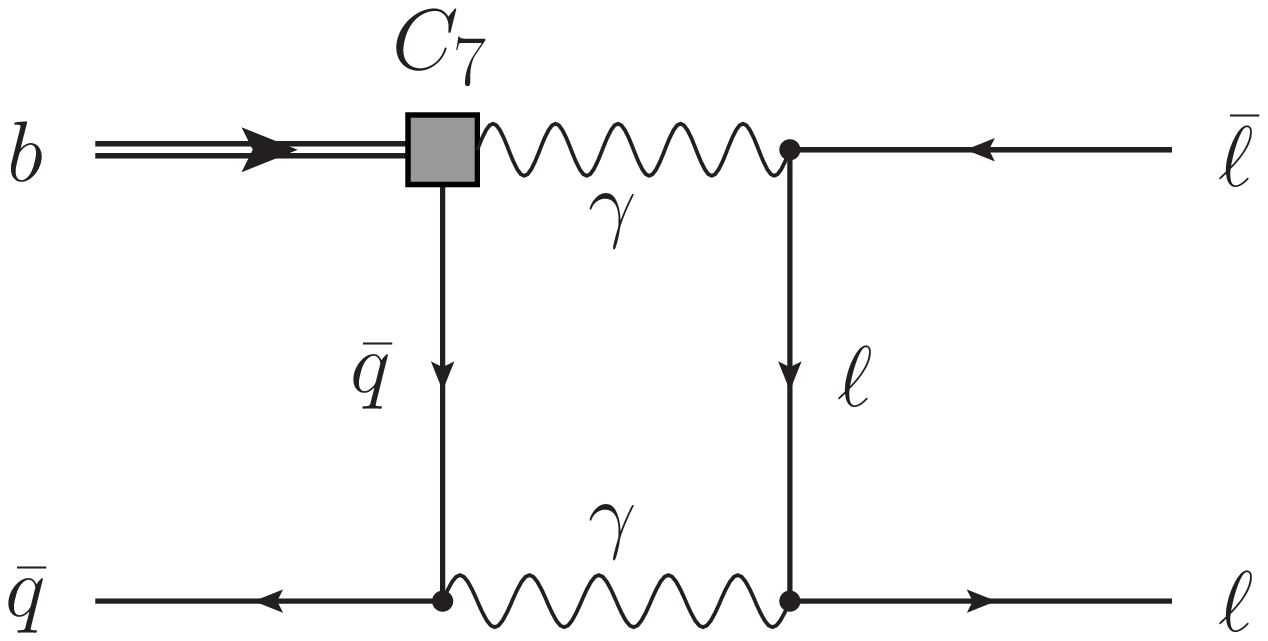}
\\[0.5cm]
\includegraphics[width=0.25\textwidth]{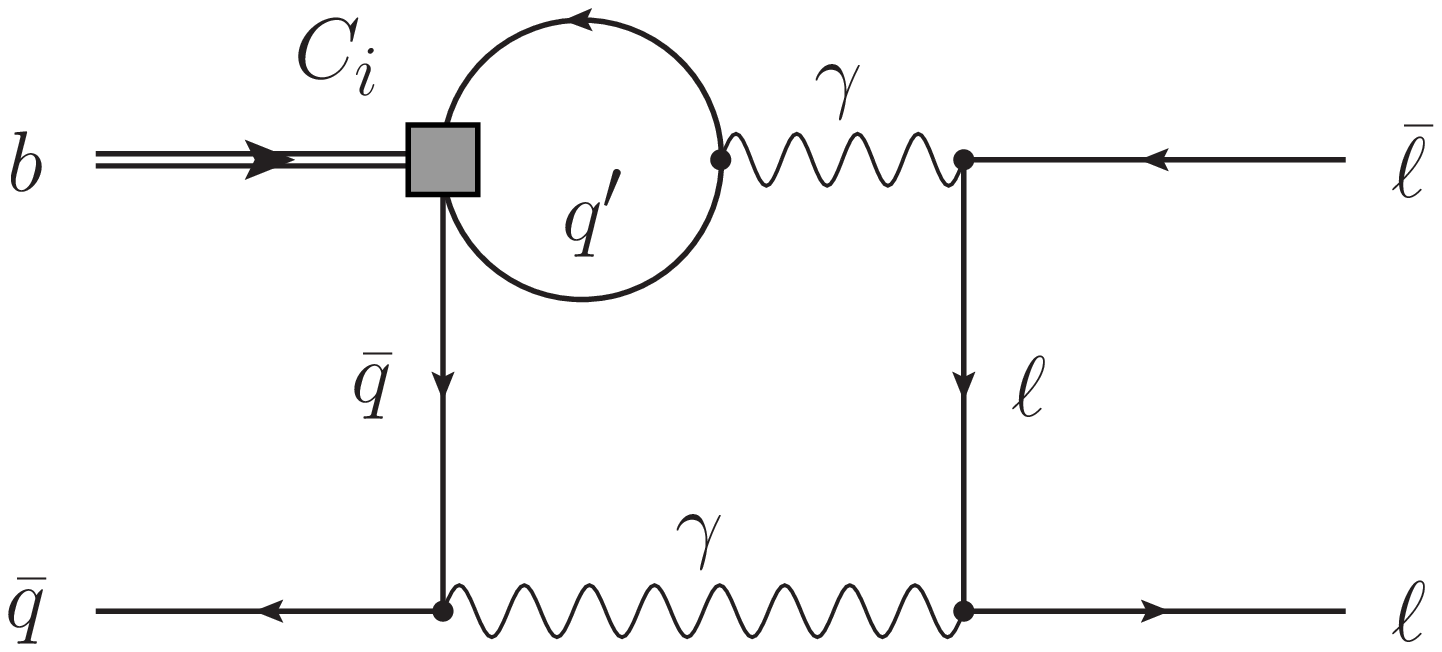}
\end{center}
\vspace*{-0.2cm}
\caption{\small Feynman diagrams that contain the power-enhanced 
electromagnetic correction. Symmetric diagrams with order of vertices on the 
leptonic line interchanged are not displayed.}
\label{fig:diagrams}
\end{figure}

We then find that the leading-order $\bar{B}_q\to \ell^+\ell^-$ decay 
amplitude plus the electromagnetic correction can be represented as 
\begin{widetext}
\vskip-0.5cm
\begin{eqnarray}
i \mathcal{A} &=& m_\ell f_{B_q} {\cal N}\,C_{10} \,\bar{\ell} \gamma_5 \ell
\nonumber\\
&& +\,\frac{\alpha_{\rm em}}{4\pi} Q_\ell Q_q\, 
m_\ell m_B f_{B_q} {\cal N}\,\bar{\ell} (1+\gamma_5) \ell
\times\Bigg\{
\int_0^1 du \,(1-u)\,C_9^{\rm eff}(u m_b^2)\,
\int_0^\infty\frac{d\omega}{\omega}\,\phi_{B+}(\omega) 
\left[\ln\frac{m_b\omega}{m_\ell^2}+\ln\frac{u}{1-u}\right]
\nonumber\\
&& -\,Q_\ell  C_7^{\rm eff} 
\int_0^\infty\frac{d\omega}{\omega}\,\phi_{B+}(\omega) 
\left[
\ln^2\frac{m_b\omega}{m_\ell^2}
-2\ln\frac{m_b\omega}{m_\ell^2}+\frac{2\pi^2}{3}
\right]
\Bigg\}+\ldots\,,
\label{eq:mainresult}
\end{eqnarray}
\end{widetext}
where the overall factor
\begin{equation}
\mathcal{N} = V_{tb} V^*_{tq} \frac{4 G_F}{\sqrt{2}}
\frac{\alpha_{\rm em}}{4 \pi}
\end{equation}
contains CKM quark-mixing elements, the Fermi constant $G_F$, 
and $Q_\ell=-1$, $Q_q=-1/3$ denote the lepton 
and quark electric charge, respectively. We use the short-hands $\bar{\ell}=
\bar{u}(p_{\ell^-})$, $\ell=v(p_{\ell^+})$ for the external 
lepton spinors. Omitted terms are power-suppressed. The two terms in 
the electromagnetic correction in the above equation arise from the 
four-fermion operator 
$Q_9 = \frac{\alpha_{\rm em}}{4 \pi} 
(\bar{q} \gamma^\mu P_L b)(\bar{\ell} \gamma_\mu\ell)$ 
and the electric dipole operator $Q_7$ in the effective weak 
interaction Lagrangian
\begin{equation}
  \label{eq:eff:Lagr}
  {\cal L}_{\Delta B=1} 
  =  \frac{4 G_F}{\sqrt{2}}\,\sum_{i=1}^{10} C_i Q_i + \mbox{h.c.} \,,
\end{equation}
with the effective operators $Q_i$  as defined in 
Ref.~\cite{Chetyrkin:1996vx}. The effective short-distance 
coefficients~\cite{Bobeth:1999mk,Beneke:2001at} 
\begin{eqnarray}
&&C_7^{\rm eff} = C_7-\frac{C_3}{3}-\frac{4 C_4}{9}-\frac{20 C_5}{3}
-\frac{80 C_6}{9}
\\[0.1cm]
&&C_9^{\rm eff}(q^2) = C_9+Y(q^2)
\end{eqnarray}
account for the quark-loop induced contributions. The relevant Feynman 
diagrams are shown in Fig.~\ref{fig:diagrams}.

An important observation on Eq.~(\ref{eq:mainresult}) is that the 
non-perturbative strong-interaction physics is no longer contained 
in the $B$-meson decay constant $f_{B_q}$ alone. Rather, the exchange 
of an energetic  photon between the lepton pair and the spectator antiquark 
$\bar q$ probes correlations between the constituents in the $B$ 
meson separated at large but light-like distances. The corresponding 
strong-interaction physics is parameterized by the inverse moment 
of the $B$-meson light-cone distribution amplitude (LCDA) $\lambda_B$, 
introduced in Ref.~\cite{Beneke:1999br},
\begin{eqnarray}
  &&\frac{1}{\lambda_B(\mu)} 
  \equiv \int_0^\infty \frac{d\omega}{\omega} \,\phi_{B+}(\omega, \mu),
\\[0.2cm]
&&  \frac{\sigma_n(\mu)}{\lambda_B(\mu)} 
  \equiv  \int_0^\infty \frac{d\omega}{\omega} 
         \ln^n\frac{\mu_0}{\omega} \,\phi_{B+}(\omega, \mu)
\end{eqnarray}
and the first two inverse-logarithmic moments, which we define as in 
Ref.~\cite{Beneke:2011nf} with fixed $\mu_0 = 1$~GeV. These parameters 
have frequently appeared in other exclusive $B$-meson decays. In the 
numerical analysis below we shall adopt \cite{Beneke:2011nf}
$\lambda_B(1~\mbox{GeV}) = (275\pm 75)~\mbox{MeV}$,
$\sigma_1(1~\mbox{GeV}) = 1.5\pm 1$, and $\sigma_2(1~\mbox{GeV}) = 3\pm 2$.
The non-locality of $\bar{q} b$ annihilation due to the photon interaction 
removes a suppression factor of the local annihilation process. The 
enhancement of the electromagnetic correction by a factor 
$m_B/\Lambda_{\rm QCD}$ in Eq.~(\ref{eq:mainresult}) arises 
from 
\begin{equation}
m_B \int_0^\infty\frac{d\omega}{\omega}\,\phi_{B+}(\omega) 
\,\ln^k\omega 
\sim \frac{m_B}{\lambda_B}\times \sigma_k\,.
\end{equation}
There is a further single-logarithmic enhancement 
of order $\ln m_b\Lambda_{\rm QCD}/m_\mu^2\sim 5$ for the $C_9^{\rm eff}$ 
term, and even a double-logarithmic enhancement of the $C_7^{\rm eff}$ 
term.

We obtained Eq.~(\ref{eq:mainresult}) in two different ways. First, from 
a standard computation of QED corrections to the four-point amplitude 
with two external lepton lines, one heavy-quark and one light-quark line, and 
second, from a method-of-region computation~\cite{Beneke:1997zp} in the 
framework of soft-collinear effective 
theory (SCET)~\cite{Bauer:2000yr,Beneke:2002ph}. The second method is 
instructive as it reveals the origin of the enhancement from the 
hard-collinear virtuality $\mathcal{O}(m_b\Lambda_{\rm QCD})$ 
of the spectator-quark propagator. A further single-logarithmic 
enhancement arises from the contribution of both hard-collinear 
and collinear (virtuality $\Lambda_{\rm QCD}^2\sim m_\ell^2$) photon 
and lepton virtuality. The double logarithm in the $C_7^{\rm eff}$ 
term is caused by an endpoint-singularity as $u\to 0$ in the 
hard-collinear and collinear convolution integral for the box diagrams, 
whereby the hard photon from the electromagnetic dipole operator 
becomes hard-collinear. The singularity is cancelled by a 
soft contribution, where the 
leptons in the final state interact with each other through the 
exchange of a soft lepton.  The relevance of soft-fermion exchange is 
interesting by itself since it is beyond the standard analysis of 
logarithmically enhanced terms in QED. We shall therefore return to a 
full analysis within SCET in a detailed separate paper.

We now proceed to the numerical evaluation of the power-enhanced QED 
correction. Let us denote $m_B$ times the curly bracket in 
Eq.~(\ref{eq:mainresult}) by $\Delta_{\rm QED}$. Since the scalar 
$\bar \ell\ell$ term in the amplitude 
$\mathcal{A}$ does not interfere with the pseudoscalar tree-level  
amplitude, the QED correction can be included in the 
expression for the tree-level $B_s \to \ell^+\ell^-$ 
branching fraction~\footnote{The given expression 
refers to the ``instantaneous'' branching fraction at $t=0$, which 
differs from the untagged time-integrated branching fraction (\ref{eq:SM:Br})
by the factor $(1-y_s^2)/
(1+y_s\mathcal{A}_{\Delta \Gamma})$~\cite{DeBruyn:2012wk}, where $y_s$ is 
related to the lifetime difference of the two $B_s$ mass eigenstates.},
\begin{equation}
\label{eq:SMtree}
\frac{\tau_{B_q}m_{B_q}^3 f_{B_q}^2}{8\pi}\,|\mathcal{N}|^2\,
\frac{m_\ell^2}{m_{B_q}^2}
\sqrt{1-\frac{4m_\ell^2}{m_{B_q}^2}}\,|C_{10}|^2\,,
\end{equation}
by the substitution
\begin{equation}
C_{10} \to C_{10} + 
\frac{\alpha_{\rm em}}{4\pi} Q_\ell Q_q \Delta_{\rm QED}\,.
\end{equation}
We calculate the Wilson coefficients $C_i(\mu_b)$ entering $\Delta_{\rm QED}$ 
at the scale $\mu_b=5\,$GeV at next-to-next-to-leading logarithmic accuracy 
in the renormalization-group evolution from the electroweak 
scale, evaluate the convolution integrals in Eq.~(\ref{eq:mainresult}) 
with $m_b=4.8\,$GeV, 
and express them in terms of $\lambda_B(1\,\mbox{GeV})$, 
$\sigma_1(1\,\mbox{GeV})$, $\sigma_2(1\,\mbox{GeV})$ specified above. 
We then find 
\begin{equation}
\Delta_{\rm QED} = (33 - 119) + i\,(9-23)\qquad (\ell=\mu)\,,
\end{equation}
where the large range is entirely due to the independent variation 
of the poorly known parameters of the $B$-meson LCDA.  
In this result the total effect is reduced by a factor of three 
by a cancellation between the $C_9^{\rm eff}(q^2)$ and  $C_7^{\rm eff}$ 
term. With $C_{10} = -4.198$, this results in a $(0.3-1.1)\%$ reduction  
of the muonic $B_s \to \ell^+\ell^-$ branching fraction. We 
update the SM prediction to 
\begin{eqnarray}
\label{eq:SMBrnew}
\overline{{\cal B}}(B_s \to \mu^+\mu^-)_{\rm SM} 
&=& (3.57 \pm 0.17) \cdot 10^{-9}\,,
\quad
\end{eqnarray}
which supersedes the one from Eq.~(\ref{eq:SM:Br}). 
To obtain this result we proceeded as in Ref.~\cite{Bobeth:2013uxa} and used  
the same numerical input except for updated values of the strong 
coupling $\alpha_s^{(5)}(m_Z) = 0.1181(11)$
and $1/\Gamma_H^{s} = 1.609(10)$~ps \cite{Patrignani:2016xqp}, 
$f_{B_s} = 228.4(3.7)$~MeV
($N_f=2+1$) \cite{Aoki:2016frl}, $|V_{tb}^* V_{ts}/V_{cb}| = 0.982(1)$ 
\cite{Bona:2016dys} and the inclusive determination of
$|V_{cb}|=0.04200(64)$ \cite{Gambino:2016jkc}. The parametric 
($\pm 0.167$) and non-parametric non-QED ($\pm 0.043$) uncertainty and 
the uncertainty from the QED correction (${}^{+0.022}_{-0.030}$) have been 
added in quadrature. Quite surprisingly, the QED uncertainty (which itself 
is almost exclusively parametric, from the $B$-meson LCDA) is now 
almost as large as the non-parametric non-QED uncertainty. 

The generation of a scalar $\bar{\ell}\ell$ amplitude in 
Eq.~(\ref{eq:mainresult}) leads to further interesting effects. 
The time-dependent rate asymmetry for $B_s$ decay into a 
muon pair $\mu^+_\lambda\mu^-_\lambda$ in the $\lambda=L,R$ 
helicity configuration is given by
\begin{eqnarray}
&& \frac{\Gamma(B_s(t)\to \mu^+_\lambda\mu^-_\lambda)- 
\Gamma(\bar B_s(t)\to \mu^+_\lambda\mu^-_\lambda)}
{\Gamma(B_s(t)\to \mu^+_\lambda\mu^-_\lambda)+ 
\Gamma(\bar B_s(t)\to \mu^+_\lambda\mu^-_\lambda)}
\nonumber\\
&&=\,\frac{C_\lambda \cos(\Delta M_{B_s}t)+S_\lambda \sin(\Delta M_{B_s}t)}
{\cosh(y_s t/\tau_{B_s})+\mathcal{A}^\lambda_{\Delta \Gamma}
\sinh(y_s t/\tau_{B_s}) }\,,
\end{eqnarray}
where all quantities are defined in Ref.~\cite{DeBruyn:2012wk}.
For example, the mass-eigenstate rate asymmetry 
$A_{\Delta\Gamma}^\lambda$  equals 
exactly $+1$, if only a pseudo-scalar amplitude exists, and is 
therefore assumed to be very sensitive to new flavour-changing 
interactions, with essentially no uncertainty from SM background. 
We now see that the SM itself generates a small ``contamination'' 
of the observable, given by 
\begin{eqnarray}
\mathcal{A}^\lambda_{\Delta\Gamma} &=& 
1-r^2 |\Delta_{\rm QED}|^2 \approx 1- 1.0\cdot 10^{-5}\,, 
\\[0.2cm]
C_\lambda&=&-\eta_\lambda\,2r \,\mbox{Re}(\Delta_{\rm QED}) 
\approx \eta_\lambda\,0.6\%\,,
\\[0.2cm]
S_\lambda&=&2r \,\mbox{Im}(\Delta_{\rm QED}) 
\approx -0.1\%\,,
\end{eqnarray}
where $r\equiv \frac{\alpha_{\rm em}}{4\pi} \frac{Q_\ell Q_q}{C_{10}}$ 
and $\eta_{L/R} = \pm 1$.
Present measurements \cite{Aaij:2017vad} set only very weak 
constraints on the deviations of $A_{\Delta\Gamma}^\lambda$ 
from unity, and $C_\lambda$, $S_\lambda$ have not yet been measured, 
but the uncertainty in the $B$-meson LCDA is in 
principle a limiting factor for the precision with which New Physics 
can be constrained from these observables.

The power-enhanced QED correction reported here may appear also relevant  
to the leptonic charged $B$-meson decay $B^+\to\ell^+\nu_\ell$, but cancels 
due to the V--A nature of the charged current. While we discussed 
only the case $\ell=\mu$ above, the other leptonic final states 
$\ell=e,\tau$ are also of interest. However, whereas the muon mass 
is numerically of the order of the strong interaction scale, the much 
larger mass of the tau lepton, and the much smaller electron mass 
imply that the results are not exactly the same. We therefore conclude 
that the systematic study of hitherto neglected electromagnetic 
corrections to exclusive $B$ decays reveals an unexpectedly complex  
structure. Its further phenomenological and theoretical implications are 
currently under investigation.\\[0.0cm] 


\noindent We thank H.~Patel for helpful communication on Package-X 
\cite{Patel:2016fam}. This work is supported by the DFG 
Son\-der\-for\-schungs\-bereich/Trans\-regio~110 
``Symmetries and the Emergence of Structure in QCD''. 



\end{document}